\documentclass[11pt,fleqn]{article}
\usepackage{appendix}
\usepackage{setspace}
\usepackage{ccaption}
\usepackage{latexsym}
\usepackage{appendix}
\usepackage{amsmath}
\usepackage{natbib}
\usepackage[none]{hyphenat}
\usepackage[bookmarks=false,colorlinks,citecolor=black,filecolor=black,linkcolor=black,urlcolor=black]{hyperref}
\hypersetup{pdftex}
\usepackage[T1]{fontenc}
\usepackage{anysize}
\marginsize{3 cm}{3 cm}{2 cm}{2.5 cm}
\usepackage{rotating}
\usepackage{url}
\usepackage{dcolumn,longtable}

\begin{document}

\title{\LARGE{Measuring market liquidity:} {\LARGE{An introductory survey}}}

\author{Alexandros Gabrielsen, Massimiliano Marzo and Paolo Zagaglia\footnote{Gabrielsen: Cass Business School; A.Gabrielsen@city.ac.uk. Marzo: Department of Economics, Universit\`{a} di Bologna; massimiliano.marzo@unibo.it. Zagaglia: Department of Economics, Universit\`{a} di Bologna; paolo.zagaglia@unibo.it. }}

\date{{\small This version: \today}}

\maketitle

\thispagestyle{empty}

\singlespacing

\begin{abstract}
Asset liquidity in modern financial markets is a key but elusive concept. A market is often said to be liquid when the prevailing structure of transactions provides a prompt and secure link between the demand and supply of assets, thus delivering low costs of transaction. Providing a rigorous and empirically relevant definition of market liquidity has, however, provided to be a difficult task. This paper provides a critical review of the frameworks currently available for modelling and estimating the market liquidity of assets. We consider definitions that stress the role of the bid-ask spread and the estimation of its components that arise from alternative sources of market friction. In this case, intra-daily measures of liquidity appear relevant for capturing the core features of a market, and for their ability to describe the arrival of new information to market participants. 
\\ \\
JEL Classification No.: G1, G10, G12. 
\end{abstract}

\newpage

\onehalfspacing

\section{Introduction}

The scope of this paper is to present an exhaustive discussion on 
the various measures for asset liquidity proposed in the literature on market
microstructure. Given the large number of liquidity measures and
methodologies employed both by practitioners and academic researchers,
this paper reviews the role of each liquidity measure by
looking at the logic behind their construction, and how they relate to each other.

Liquidity is often pointed at as a key concept in financial markets. It is a very
elusive one though. In general terms, the concept of liquidity often denotes a desirable function that should reflect a well organized financial market. A market is often said to be
liquid when the prevailing structure of transactions provides a prompt and secure link between the demand and supply of assets, thus delivering low transaction costs. Providing a rigorous definition of market liquidity has, however, proven to be a cumbersome task. 

Differently from a widely-quoted contribution of \citet{bak}, this paper considers definitions of market liquidity that emphasize the role of the bid-ask spread and the estimation of its components. The difference between bid and ask quotes for an asset provides a liquidity measure applicable to a dealer market, rather than a broker market. Despite this, it is possible to compute approximations that mimic the difference between bid and ask quotes even in broker markets. Hence, the role of intradaily measures of liquidity can capture the core features of a market, such as the arrival of new information to the trading parties. 

The first major task of any study on liquidity consists in providing an exhaustive definition. Laying down this concept properly involves the specification of two additional concepts, namely the the transaction time - i.e. the speed of executing transactions -  and the pure transaction costs - i.e. the price paid by investors for the liquidity services. 

The time of transaction is related to the demand pressure generated by the public. This takes the form of a request for a quick execution of the order placed in the market. At the same time, an order request involves the opportunity for the investor to buy or sell an asset at the prevailing price, or at a price close to the one prevailing in the market. 

These intuitive considerations lay down the ground for a relevant concept of liquidity. In other words, an asset is liquid if it can be quickly exchanged at a minimal cost. A similar definition can be applied also to an asset market as a whole. In this sense, a market is liquid if it is possible to buy and sell assets at a minimal cost without too much delay from order placement. 

Another important aspect concerns the extent to which asset prices are affected by the trading activity. In a liquid market, price variations should not to be determined by transaction costs. In other words, block size transactions should have a minimal impact on prices. Prices usually change both in anticipation and in response to order flows. Hence, it becomes crucial to understand the extent to which the amount of transactions or order size can determine large swings. In a thin market, prices are highly responsive to trade size. In a liquid or deep market instead, prices can be affected by order flows only to a minor extent.

Insofar as thera are different concepts of asset liquidity, different measures of liquidity focus on alternative aspects of the measurement problem. Some measures concentrate on the role of the volume size. Other indices are related to the execution-cost aspect of liquidity. The indices based on volume information are related to the price impact of transactions. When properly aggregated, they provide also synthetic measures of the liquidity present in an entire market. On the other hand, the indices based on execution costs are meant to evaluate the properties of an asset by looking at the cost paid to the market maker (dealer or specialist)  for matching the demand and the supply. These analyses are generally based on the bid-ask spread and its variations. In fact, when a dealer or a specialist revises the bid-ask quotes, a careful study of bid-ask spread components can reveal information on the sources of illiquidity. 

The literature identifies three main component of the bid-ask spread. These arise from order processing, adverse information and inventory costs. A high level of competition between intermediaries allows for a reduction of the order processing component and improves the liquidity condition of the market. The informational component of the bid-ask spread sheds light on the degree of efficiency due to the presence of hidden information or insider trading.\footnote{These considerations suggest that a careful analysis of liquidity is a crucial step towards the design of a proper regulatory activity for both exchanging parties and the intermediaries.}

The content of the paper is organized as follows. Section 2 discusses the various concepts of liquidity with a view on their implications for asset pricing. Section 3 reviews the measures of liquidity based on information from traded volumes. Section 4 considers the indices computed from asset prices. Section 5 focuses on the role of transaction costs as a source of asset  illiquidity. 

\section{Why is market liquidity important?}

The relevance of market liquidity arises from its connection with the institutional organization of a market. Both aspects tend to influence each other and produce effects on the efficiency of market transactions, as documented by \citet{am88,am91}. 

As a starting point, let us consider the characteristics of a liquid market. According to \citet{bak96}, we can identify three main properties:
\begin{enumerate}
\item Depth: a market is deep when there are orders both above
and below the trading price of an asset.

\item Breadth: a market is broad when there is a large volume of
buying and selling orders. The spread is large when the order flow is
scarce.

\item Resiliency: a market is resilient if there are many orders
in response to price changes. There is a lack of resiliency when the order flow does
not adjust quickly in response to price swings.
\end{enumerate}
All these aspects play a crucial role in the evaluation of the
structure of a financial market. In fact, the availability of liquidity has
important consequences both on the prices of assets, and on the degree of competition between market actors. 

Abstract properties may not allow to provide an operational definition of liquidity which can be confronted with the data. Different sources provide definitions of liquidity that are not fully satisfactory because they stress a specific aspects different from one another. According
to John Maynard Keynes, an asset is liquid if 
\begin{center}
\begin{quote}
{\it "it is more certainly realizable at
short notice without loss." }
\end{quote}
\end{center}
This quotation highlights two aspects, namely the riskiness of the realizationof an asset value, and the presence of a marketplace where negotiations can take place without adverse price oscillations. 

Subsequent contributions have pointed out the role of speed and the costs associated to market exchanges. For example, Massimb and Phelps (1994) focus on the importance of immediacy. Liquidity can be defined as the 
\begin{center}
\begin{quote}
{\it "market ability to provide immediate execution for an
incoming market order (often called "immediacy") and the ability to
execute small market orders without large changes in the market price(often
called "market depth" of "resiliency")." }
\end{quote}
\end{center}

The core point of the concept of liquidity is the possibility to exchange a given asset in the market without dramatic changes in the prevailing market price. Sensible empirical implementations of this idea are hard to construct because the `true' degree of liquidity is unobservable. In particular, this is well represented by the difference between the observed
transaction price and the price that would occurred in complete absence of transaction costs.

\subsection{A selected discussion on asset-pricing implications}

As stressed in a long series of papers, market liquidity has important asset pricing implications.\footnote{For instance, see the contributions of \citet{ami,am87,am91}.} The rate of return on a given asset should include the compensation to investors for potential losses arising from transaction costs. The presence of a liquidity cost of one percent affects an asset price by more than one percent because of the repeated trades. Thus, with higher
transaction costs, the market will experience lower asset prices and higher rates of return. An illiquid asset will offer a higher rate of return in order to
compensate the investor for bearing liquidity cost at any different date. 

This proposition can be understood in the following way. The present discounted value of transaction costs of a liquid asset traded frequently is higher than the one of a less liquid asset, which is thus traded less frequently. Thus, the return of the more liquid asset will be higher than
the return on the less liquid, because of the distorting effect due to
transaction costs. 

All these considerations should represent a concern relevant for the analyses carried out by portfolio managers. \citet{am86,am87,am88,am91} propose a model to evaluate the asset pricing implications of market liquidity. This is based on the assumption that asset returns are an increasing and concave function of the spread. The idea behind this framework can be thought of as a 'clientele effect'. This is the tendency of investors with longer holding periods to select assets with higher spreads, so that expected returns net of trading costs increase with the holding period. 
In this case, higher spreads due to the presence of higher transaction costs will yield higher net returns. As a result, an investor with a long intertemporal horizon will gain by investing in assets characterized by higher spreads.

The prediction offered by the model by \citet{am86} can be
tested by estimating the following regression for a portfolio $j$ of assets:
\begin{equation}
R_{t}\left( j\right) =c+\alpha \beta _{t}\left( j\right) +\gamma \log
S_{t}\left( j\right)  \label{e1}
\end{equation}
In the original framework, $R_{t}\left( j\right) $ denotes the average monthly rate of return on a
stock included in the portfolio j in excess of the 90-day return on Treasury
bonds, $\beta _{t}\left( j\right) $ is the beta coefficient for portfolio j,
while $S_{t}\left( j\right) $ is the average bid-ask spread. 

The empirical analysis based on estimates for \ref{e1} show a high level of significance for all the arguments of regression. In particular, \citet{am88,am91} show that average portfolio returns increase with the spread, and the spread effect persists if firm size is included in equation  \ref{e1} as an additional regressor. 

The importance of liquidity for portfolio management is also documented by
the role of technical analysis as indicator for price pressure. The academic profession tends to consider technical analysis as partly irrelevant for asset pricing. However, typical technical analysis indicators such as the traded volume can turn out extremely useful. As shown by \citet{beoh},  information on traded volumes provides important insights that are usually not conveyed by simple price statistics.\footnote{The influence of volume on returns is also analyzed both empirically and theoretically by Campbell, Grossman and Wang (1993).}

Other authors have documented several anomalies or puzzles that link volume or trade indicators to liquidity. For example, a relationship between returns and volume is documented in the literature on weekends effects, started by \citet{frenchroll}, and in the contributions on intra-day patterns described by \citet{mw}. Joint indicators of liquidity and volume are also often employed in the pricing of infrequently traded stocks \citep[e.g., see][]{beoh}.

Overall, this strand of research suggests that information is an important pricing factors in asset markets. Information is often reflected in the frequency of transactions and, as such, in market liquidity. We should stress that these aspects concern the demand side of a market, namely the pricing pressure determined by purchasing activity. The remainder of the paper focuses on liquidity as a supply-side factor.

\section{Volume-based liquidity measures}

In this section we present the liquidity indices proposed in the early stages of the market microstructure literature. Their emphasis is on the relationship between price and quantity of an asset. These measures evaluate the degree of price impact of a transaction of a specific size.

\subsection{Trading volume}

A rough measure of liquidity is represented by the traded volume. This consists in the
amount exhanged between market actors in buying and selling activities for a single asset or for the market as a whole. Some researchers consider trading volume as an inappropriate liquidity index, though. The reason lies in the issue of double counting involved. A transaction on the buy side can be also recorded as transaction on the seller side. A more suitable measure is provided by the ratio between trading volume and market capitalization. 

According to recent contributions, the trading volume of an asset is one of the key determinant for the whole pricing structure. For example, \citet{beoh} show that the volume traded generates information that cannot be extracted from alternative statistics. Because of widespread availability of data, trade volume represents a sort of a preliminary step
towards a more complete analysis of market liquidity.

\subsection{The conventional liquidity ratio}

The liquidity ratio, also called `conventional liquidity ratio', is probably one of the liquidity measures most frequently applied in the empirical analysis. This index provides a measure for how much traded volume is necessary to induce a  price change of one percent. Volumes and prices are the key ingredients. The analytical expression of the liquidity ratio for asset $i$ is:
\begin{equation}
LR_{it}=\frac{\sum\limits_{t=1}^{T}P_{it}V_{it}}{\sum\limits_{t=1}^{T}\left|
PC_{it}\right| }  \label{lr}
\end{equation}
where $P_{it}$ is the price of asset $i$ on day $t$, $V_{it}$ denotes the
volume traded, and $\left| PC_{it}\right| $ is the
absolute percentage price change over a fixed time interval,
given by $PC_{it}=P_{it}-P_{it-1}$. 

The liquidity ratio is usually computed for a number of assets and is aggregated over a  pool with similar characteristics. The time interval $(T,t)$ adopted to compute the
index is typically chosen arbitrarily. However, the index is often
calculated over a monthly time scale, so that the numerator denotes the total
volume of the traded assets over the previous four
weeks. Instead, the numerator is the absolute value of the daily percentage price
changes of the stock over the last four weeks. The higher the ratio $%
LR_{it} $ is, the higher the liquidity of asset $i$ will be. This means that
large volumes of trades have little influence on price. Obviously, this conceptual framework focuses more on the price aspect than on the issue of time or on the execution costs typically present in a market.

\subsection{The index of \citet{martin}}

\citet{martin} proposes a liquidity index where a stationary distribution of
price changes is assumed to hold through the entire transaction time. 
The analytical expression for the index takes the form:
\begin{equation}
MLI_{t}=\sum_{i=1}^{N}\frac{\left( P_{it}-P_{it-1}\right) ^{2}}{V_{it}}
\label{martin}
\end{equation}
where $P_{it}$ is the closing price and $V_{it}$ denotes the traded volume. The reader should otice that the index is computed over the total number of asset for the market. $MLI_{t}$ is considered as a suitable index for the market as a whole, while the liquidity ratio is best suited for a single asset. 

A higher value for $MLI_{t}$ implies less liquidity because of the influence of price dispersion. Another interpretation of the index is the following. The higher the ratio, the higher the price dispersion relative to the traded volume, and the lower is the liquidity of the market. In fact, prices appear to be uncorrected with trading volume. As such, they may reflect only changes of information or events not necessarily related with the trading process.

For its characteristics, Martin's (1975) liquidity index produces meaningful results if computed on a daily basis. To obtain sensible outcomes for longer time horizons, one needs to compute a weighted average of several indices derived for shorter time intervals.

\subsection{The liquidity ratio of \citet{huihe}}

\citet{huihe} introduce an additional index that measures the
liquidity of a single asset. As such, it cannot be directly employed for the
market as a whole without using appropriate aggregation techniques. In practice, this index constructs a metric between the largest price change divided by the ratio of traded volume to
market capitalization. 

In what follows, we drop the superscipt $i$ from the formula for reasons of notational convenience. The mathematical expression of the index is:
\begin{equation}
LR_{HH}=\frac{\left( P_{\max }-P_{\min }\right) /P_{\min }}{V/\left( S\cdot
\overline{P}\right) }  \label{hh}
\end{equation}
where $P_{\max }$ is the highest daily price over a 5-day period,
$P_{\min }$ is the lowest daily price over the same horizon, $V$ is the
total volume of assets traded over a 5-day period, $S$ is the total number
of assets outstanding and $\overline{P}$ denotes the average closing price. A higher value for the index $LR_{HH}$ implies lower liquidity. 

A quick inspection of equation (\ref{hh}) reveals that the logic behind the
construction of this index is not very different from that underlying $%
MLI_{t}$. In fact, the denominator of (\ref{hh}) is the traded volume
adjusted for market capitalization and the numerator indicates the widest
percentage price change over a 5-day horizon. 

According to the existing literature, the ratio proposed by \citet{huihe} suffers at least
of two shortcomings. First, the time period consisting of 5 days is arguably 
too long for the index to detect market anomalies, given the fact that asset prices can quickly adjust to liquidity problems. The second critical point is related to the choice of variables. For instance, if we focus on stocks quoted in a dealer market, such as the NASDAQ, high-quality price data may not be readily available. In this case, it is possible to replace $P_{\max }$ and $P_{\min }$ with the bid-ask spread. However, this represents a problematic approach because the bid-ask spread quotes are often less volatile than prices. Hence, the use of bid-ask quotes may bias downward the analysis of liquidity. This issue motives the adoption of an alternative liquidity measure.\footnote{To deal with so-called `company ratio problem', 
\citet{huihe} normalize the liquidity ratio by the value of
outstanding shares.}

\subsection{The turnover ratio}

The turnover ratio $TR_{t}^{i}$ for an $i$ at time $t$ is usually defined
as follows:
\begin{equation}
TR_{t}^{i}=\frac{Sh_{t}^{i}}{NSh_{t}^{i}}  \label{tur}
\end{equation}
where $Sh_{t}^{i}$ is the number of asset units traded at time t for stock i, and
$NSh_{t}^{i}$ is the total number of asset units outstanding. The
index proposed in (\ref{tur}) is computedfor a single time period, which could be
a day or a month. Often it is used to compute an average over a prespecified sample period as:
\begin{equation}
TR_{T}^{i}=\frac{1}{N_{T}}\sum_{t=1}^{N_{T}}TR_{t}^{i}  \label{turag}
\end{equation}
with a number of sub-periods $N_{T}$. Thus, in this expression, we compute the mean of the turnover ratio over a defined sample period. 

The indices outlined earlier can be included into parametric models for asset prices. For instance, we can consider a regression of asset returns to test for the statistical significance of various liquidity measures. According to\citet{am86}, turnover is negatively related to the illiquidity costs of stocks. 

\citet{datar} propose a test for the role of liquidity that is different from the one proposed by \citet{am86}. In particular, they use the turnover
rate as a proxy for liquidity. This test can be widely employed because of
its simplicity and data availability. From \citet{am86}, in
equilibrium, liquidity is correlated with trading frequency. Therefore, by
directly observing the turnover rate, it is possible to obtain the latter as a
proxy for liquidity. \citet{datar} perform the
following regression in cross sectional data:
\[
R_{t}^{i}=k_{0}+k_{1}TR_{t}^{i}+k_{2}b_{t}^{i}+k_{3}lsize_{t-1}^{i}+k_{4}%
\beta _{t}^{i}+e_{t}
\]
where $R_{t}^{i}$ is the return of stock i at month t, $TR_{t}^{i}$ is the turnover ratio at month t, $b_{t}^{i}$ is the book to market ratio expressed as the natural logarithm of book value to market value for each individual firm, $lsize_{t}^{i}$ is the natural logarithm of total market capitalization of firm i at the end of the prior month. Finally, $\beta_{t}^{i}$ is the coefficient for the i-th stock computed for stocks belonging to a portfolio of homogeneous stocks and $e_{t}$ are the residuals from the estimation of the above equation. The results suggest that stock returns are a
decreasing function of the turnover rates. This relation is robust after controlling for $b_{t}^{i}$, $lsize_{t-1}^{i}$, and $\beta _{t}^{i}$.

\subsection{The market-adjusted liquidity index}

\citet{huihe} propose a measure for liquidity that takes into account the systematic sources for risk. The construction of the market index involves two steps. In the first step, a market model for the asset return is estimated to control for the effects of average market conditions on price changes. For stock prices, this stage typically consists in estimating the following equation:
\begin{equation}
R_{it}=\alpha +\beta R_{mt}+\varepsilon _{it}  \label{hh94}
\end{equation}
where $R_{it}$ is the daily return on the $i$-th stock, $R_{mt}$ is the daily market return on the aggregate stock market index, $\alpha $ is a constant, $\beta$ measures the systematic risk, and $\varepsilon _{it}$ denotes a measure of idiosyncratic risk. 

The motivation for using this model relies on the idea that part of the stock's specific risk reflects the liquidity in the market. Thus, more liquid stocks display smaller random price fluctuations, and tend to perform as the market model would suggest. In other words, larger price dispersion is a characteristics of stocks with low liquidity that deviate from the market model. 

The second step in the construction of the index consists in the definition of a model for idiosyncratic risk.  This can be formalized as:
\begin{equation}
\varepsilon _{it}^{2}=\phi _{0}+\phi _{1}\Delta V_{it}+\eta _{it}
\label{hh94b}
\end{equation}
where $\varepsilon _{it}^{2}$ are the squared residuals from equation (\ref{hh94}), $\Delta V_{it}$ is the daily percentage change in dollar volume traded, $\eta _{it}$ is an i.i.d. residual with zero mean and constant variance.

The market-adjusted liquidity ratio is identified as the coefficient $\phi _{1}$ in equation (\ref{hh94b}). A small value of $\phi _{1}$ indicates that prices change little in response to variations in volume. This measure takes into account the price effect arising from changes in
liquidity conditions, which are mimicked by the change in trading volume. A liquid stock is characterized by a low exposure to liquidity risk which is, in turn, measured by a low $\phi_1$. 

This liquidity measure provides sensisble results on the assumption that asset prices behave
according to the market model. However, if deviations from the market model are due to swings in volume, there is an identification problem. Despite this issue, the market-adjusted liquidity index provides for a simple way to test for liquidity effects. In the current literature, there is a widespread application of this index to both dealer and auction market.

\subsection{An explicit illiquidity measure}

The role of traded volume is central in the liquidity measures proposed in the recent years. An interesting index of illiquidity is introduced by \citet{amihud}:
\begin{equation}
ILLIQ_{T}^{i}=\frac{1}{D_{T}}\sum_{t=1}^{D_{T}}\frac{\left|
R_{t,T}^{i}\right| }{V_{t,T}^{i}}  \label{illiq}
\end{equation}
where $D_{T}$ is the number of days for which data are available, $R_{t,T}^{i}$ is the return on day t of year T, and $V_{t,T}^{i}$ is the daily volume. The day-$t$ impact on the price of one currency unit of volume traded is given by the ratio $\frac{\left| R_{t,T}^{i}\right| }{V_{t,T}^{i}}$. The illiquidity measure (\ref{illiq}) is the average of the daily impacts over a given sample period. 

This index is very close to the liquidity ratio. The latter provides an understanding of the link between volume and price change. The illiquidity index provides only a rough measure of the price impact. Differently from the bid-ask spread, the main advantage of this index relies on the wide availability of data for its computaiton, especially for those markets that do not report sophisticated measures of the spread.

\citet{amihud} has introduced the illiquidity index to investigate the influence of
market conditions on stock returns. His framework introduces a cross-sectional test by selecting a sample of stocks quoted on the NYSE. The testing model takes the form
\begin{eqnarray}
R_{m,T}^{i} &=&\lambda _{0}+\lambda _{1}ILLIQMA_{m,T-1}^{i}+\lambda
_{2}ATR_{m,T-1}^{i}+\lambda _{3}v_{m,T-1}^{i}+\lambda
_{4}p_{m,T-1}^{i}+\lambda _{5}c_{m,T-1}^{i}+  \nonumber \\
&&\lambda _{6}dy_{T-1}+\lambda _{7}R_{100}^{i}+\lambda
_{8}R_{T-1}^{i}+\lambda _{9}\sigma R_{T-1}^{i}+\lambda _{10}\beta
_{T-1}^{i}+u_{t}^{i}  \label{ami20}
\end{eqnarray}
The stock return $R_{m,T}^{i}$ in month m for year T is regressed over several variables, including a constant $\lambda _{0}$, the mean adjusted illiquidity measure at the end of year T-1, $ILLIQMA_{m,T-1}^{i}$, the mean adjusted turnover ratio $ATR_{m,T-1}^{i}$,
the log of traded volume, $v_{m,T-1}^{i}$, the log of stock price $p_{m,T-1}^{i}$, the log of capitalization $c_{m,T-1}^{i}$, the dividend yield $dy_{T-1}$, computed as the sum
of the annual cash dividends divided by the end-of-year price. Moreover, $R_{100}^{i}$ and $R_{T-1}^{i}$ are the cumulative stock returns over the last 100 days and the entire year, respectively, $\sigma R_{T-1}^{i}$ is the standard deviation of the stock daily return during year T-1, and $\beta_{T-1}^{i}$ is the beta of stock i computed for portfolios of stocks of homogeneous size. 

The mean-adjusted illiquidity measure takes the form:
\[
ILLIQMA_{m,T}^{i}=\frac{ILLIQ_{m,T}^{i}}{AILLIQ_{m,T}}
\]
where $AILLIQ_{m,T}$ is the cross-stock average illiquidity for the stocks included in the regression model. In general, the annual average illiquidity across stocks is defined as:
\[
AILLIQ_{T}=\frac{1}{N_{T}}\sum_{i=1}^{N_{T}}ILLIQ_{T}^{i}
\]
where $N_{T}$ is the number of stocks in year $T$. The variable $ATR_{m,T}^{i}$ is computed in the same way. This transformation allows to take into account the time-series variability in the estimated coefficients, which can arise from high volatility associated to illiquidity. 

The empirical results show that the illiquidity measure is statistically significant for NYSE stocks during the period 1964-1997. The coefficient of the illiquidity measure on stock returns has a positive sign. Stock turnover, instead, delivers a negative coefficient. The estimated parameters on illiquidity and the turnover can provide a joint measure of liquidity. These results stress the importance of liquidity for stock returns. Moreover, if stock returns are computed in excess of the Treasury bill rate, the model results show that the compensation for expected market illiquidity is still sizeable.

\subsection{General comments on volume-based measures}

We can point out at least three issues arising from the use of liquidity indices
based on volume.

First of all, these indices fail to distinguish between transitory and persistent price effect of swings in traded volume. A transitory effect can often be explained as a temporary lack of liquidity in the market, or arise from the pure transaction cost component. A permanent price effect is a price change due to the presence of informational effects because of better informed traders. This permanent effect is related to changes in the fundamental value of assets anticipated by part of the market because of inside information.

The distinction between transitory and permanent price effects can also be thought of as a problem similar to the decomposition of a time series into a stationary and a random-walk component, as studied by \citet{beveridge}. The dichotomy between permanent and non-permanent effects can be identified from pricing errors. These consist in  the difference between the `efficient' unobserved price and the actual transaction price. 

A pricing error can be decomposed into an information-related component and an uncorrelated term. The latter arises from price discreteness, temporary liquidity effects and inventory control. Information-based pricing errors are related to adverse selection. The presence of traders with superior information about assets, and a lagged adjustment of the market to new information. 

As discussed by \citet{hs},  this decomposition can be obtained only by studying the components of the bid-ask spread. \citet{frenchroll} suggest that the role of information is crucial in determining the volatility of returns. Price volatility can be the result of informational asymmetry, rather than a consequence of lack of liquidity. These are aspects that cannot be accounted for by volume-based indices. 

A second problem with volume indices is that they do no show how a sudden order arrival can affect prices. This is the so called `order-induced effect'. In other words, volume indices take into account only past links between changes in prices and volume. The reason is that these indices are not based on theoretical models of dealer/specialist behavior.

An additional issue is discussed by Marsh and Rock (1986). They argue that conventional liquidity indices tend to overestimate the impact of price changes on large transaction deals. Arguably, they also underestimate the effect of price changes on small transactions. This issue arises from the lack of proportionality between prices and volume that characterizes all the volume-based liquidity measures.

Despite these shortcomings, measures of volume can be employed fruitfully to model liquidity for agency markets rather than for dealer markets. In fact, the problem, especially for the volume indices computed on a daily basis, is that they do not take into account the effect of large block trades, which are instead very common in dealer markets. On the other hand, these measures represent a useful starting point for a more careful analysis.

\section{Price-variability indices}

In this category we can include the measures that infer asset or market liquidity directly from price behavior. We consider the \citet{mr} liquidity ratio and the variance ratio, together with its implications for market efficiency. A second group of measures infers the liquidity condition by using mere statistical techniques.

\subsection{The of liquidity ratio \citet{mr}}

Differently from the liquidity measures considered so far, \citet{mr} assume that price changes are independent from trade size, except for large traded blocks. This is based on the argument that standard liquidity ratios are strongly affected by trade size. The expression for this index is given by:
\begin{equation}
LR_{MR}^{i}=\frac{1}{M^{i}}\sum_{m=1}^{M^{i}}\left| \frac{%
P_{m}^{i}-P_{m-1}^{i}}{P_{m-1}^{i}}\right| \cdot 100  \label{mr}
\end{equation}
where $M^{i}$ is the total number of transactions for asset $i$ over a given period. The expression after the summation term denotes the absolute value of percentage price change over two subsequent periods. Intuitively, the index (\ref{mr}) considers the relation between the percentage price change and the absolute number of transactions, rather than the traded volume. In some sense, this index shifts the attention from the aggregate market to the microstructure, which is represented by number of transactions within a given time horizon. In fact, differently from the volume-based indices where traded volumes drive the scaling effect, here the scaling variable is the number of transactions. This reflects the idea that the liquidity of an asset is better represented by the price effects of transactions, rather by the impact on volumes.

To provide a better explanation, let us consider two assets. Asset A is traded in large blocks once a day, while stock B is exchanged for the same total volume but for transactions of smaller size each. Common sense would lead us to suggest that asset B is more liquid than asset A. Unfortunately, even if price changes for both A and B were similar across markets, we would not be able to reach this conclusion by looking at volume-based liquidity measures. This example helps to clarify the value generated by the liquidity index of \citet{mr}. 

The main issue with this type of index is determined by the arbitrariness involved in its formulation. In particular, the length of the period over which the index can be computed is not explicitly specified. It is clear, however, that an index computed on a hourly basis can deliver results different from those of a daily or weekly time span. Owing to its underlying properties, it is reasonable to adopt The March and Rock ratio for short horizons. Differently from alternative indices though, this measure is suitable for both dealer and auction markets.

\subsection{The variance ratio}

The variance ratio is one of the most widely-used indices in the literature. Owing to its versatility, it can be applied to contexts indirectly connected with market liquidity, such as the study of volatility and intraday effects. This liquidity measure, also called market
efficiency coefficient (MEC), measures the impact of execution costs on
price volatility over short horizons. 

The idea behind the construction of this index can be summarized as follows. With high execution costs, asset markets are characterized by price volatility in excess of the theoretical volatility of equilibrium prices. Therefore, a more liquid market implies a smaller variance of transaction prices around the equilibrium price. The reason is that the difference between actual and equilibrium price in a liquid market is smaller than what one should observe in an illiquid market.

Denote by $var\left( R_{T}^{i}\right) $ the long-term variance, and by $var\left( Z_{T}^{i}\right) $ the short-term variance of asset return $i$. Let $T$ be the number of subperiods into which longer periods of time can be divided. The variance ratio $VR^{i}$ can be defined as follows:
\begin{equation}
VR^{i}=\frac{var\left( R_{T}^{i}\right) }{T\cdot var\left( Z_{T}^{i}\right) }
\label{vr}
\end{equation}
This index proposes a metric that compares the long-term variance with the short-term variance. When $VR^{i}<1$, it suggests that the market is illiquid. In other words, the short-term retur is higher than the long-term return. If we assume that the markets are in equilibrium in the long run, this implies a large discrepancy between the short and long-term equilibrium return. Of course, when the two returns coincide, the liquidity index is equal to one. 

The variance ratio is often used to test for market efficiency. This is done by measuring the deviation of an asset price from the random hypothesis. To provide intuition on this point, let the asset price $P_{t}$ follow a random-walk process:
\begin{equation}
P_{t}=P_{t-1}+\eta _{t}  \label{pt}
\end{equation}
where $\eta _{t}$ is a homoskedastic disturbance uncorrelated over time,
i.e. $E\left( \eta _{t}\right) =0$, $Var\left( \eta _{t}\right) =\sigma
_{\eta }^{2}$, $E\left( \eta _{t}\eta _{\tau }\right) =0$ for all $\tau \neq
t$. Under the random walk hypothesis, from (\ref{pt}) we obtain $\Delta P_{t}=\eta
_{t}$. To construct the variance ratio, we can show that:
\[
var\left( P_{t}-P_{t-2}\right) =var\left( P_{t}-P_{t-1}\right) +var\left(
P_{t-1}-P_{t-2}\right) =2\sigma _{\eta }^{2}
\]
\[
var\left( P_{t}-P_{t-T}\right) =T\sigma _{\eta }^{2}
\]
Therefore, under the random walk hypothesis, $\Delta P_{t}=R_{T}$, which delivers the variance ratio:
\[
VR_{T}=\frac{var\left( P_{t}-P_{t-T}\right) }{T\sigma _{\eta }^{2}}=1
\]
For $VR_{T}=1$, there are no deviations from the random walk hypothesis. 

Now let us consider the implications of the deviation from the random walk hypothesis. Let us look at the following case:
\[
var\left( R_{t}+R_{t-1}\right) =2Var\left( R_{t}\right) +2Cov\left(
R_{t}R_{t-1}\right)
\]
The variance ratio can be constructed from:
\begin{eqnarray*}
VR\left( 2\right) &=&\frac{2Var\left( R_{t}\right) +2Cov\left(
R_{t}R_{t-1}\right) }{2Var\left( R_{t}\right) }=1+2\left[ \frac{Cov\left(
R_{t}R_{t-1}\right) }{2Var\left( R_{t}\right) }\right] = \\
&=&1+2\rho \left( 1\right)
\end{eqnarray*}
where $\rho \left( 1\right) $ is a proxy for the correlation coefficient,
whose expression is given by:
\[
\rho \left( 1\right) =\frac{Cov\left( R_{t}R_{t-1}\right) }{2Var\left(
R_{t}\right) }
\]
If we generalize this argument, we obtain a general expression for the variance ratio:
\begin{equation}
VR\left( T\right) =\frac{Var\left( R_{t}\right) }{TVar\left( R_{T}\right) }%
=1+2\sum_{s=1}^{T-1}\left( 1-\frac{s}{T}\right) \rho \left( s\right)
\label{vr2}
\end{equation}
With serially-uncorrelated asset returns, i.e. if $\rho\left( s\right) =0$, for $s>1$, the variance ratio is equal to 1. With autocorrelation of order 1, 
\[
R_{t}=\phi R_{t-1}+\varepsilon _{t}
\]
and $E\left( \varepsilon _{t}\right) =0$, $Var\left( \varepsilon
_{t}\right) =\sigma _{\varepsilon }^{2}$, the expression for the variance
ratio becomes:
\[
VR\left( T\right) =1+2\sum_{s=1}^{T-1}\left( 1-\frac{s}{T}\right) \phi ^{k}
\]

The variance ratio can be computed over arbitrary time intervals. For example, \citet{hs} calculate it over three distinct time intervals. They consider the ratio of two-day to half-hour variance, the
ratio of one-day to one-hour variance, and the ratio of two-day to one-day return variance. The logic behind this analysis lies in the different informational content of short-term and long-term variance. In fact, a sequence of short-term transactions tends to affect the market price in a way more marked than a set of transactions measured over a longer period.

From these considerations, it is reasonable to expect a value for the variance ratio index larger than unity in the presence of sequential information arrival, market-maker intervention and other factors implying undershooting of price level. It is clear, however, that this index cannot account for all the causes of liquidity costs.

The variance ratio displays two additional shortcomings. The first one is related to its sensitiveness to the time interval chosen for its calculation. In fact, this can potentially generate contrasting results when the time chosen is differently chosen. A second drawback concerns the fact that it is relation to a notion of equilibrium prices that are unobservable. The variance ratio is, however, measured from actual transaction prices. This implies that it takes into account the trading activity occurred both inside or outside the limits of the bid-ask spread.

\subsection{Event studies}

The event study methodology consists in examining asset price behavior around the time of a particular event of an informational announcement. This method is well suited to study assets around their time of issuance. This is a time when the expectation of obtaining buoyant liquidity conditions tends to generate price pressures as the asset is introduced in the market. 

Average market conditions can provide insights on liquidity. With abnormally high returns as an asset is introduced in the market, an additional supply of liquidity provides benefits for the market by generating higher returns. However, this observation can be interpreted in an alternative way. High returns can be thought of as way to compensate investors for the lack of efficient liquidity services. This inefficiency can arise from the presence of transaction costs because of the existence of transaction costs. In other words, when the discounted value of future transaction costs is incorporated in price quotations, asset returns account also for liquidity effects.

As this brief discussion suggests, it is difficult to provide a widely accepted interpretation of changes in liquidity by considering only the observed patterns of asset returns and volume exchanged. \citet{brown1,brown} and \citet{pete} suggest that there is no unique way to analyze liquidity through event studies. A general prescription is that this type of analysis should complement the information from a set of indices usually adopted for technical analysis to provide a better assessment of the event under scrutiny.

\subsection{Estimation methods based on vector autoregressions}

Vector autoregressive (VAR) models are commonly used in macroeconomics to identify the effects of various shocks on the structure of the economy. In the microstructure literature, VAR models are employed to study the transmission channels of shocks across markets. 
A representative example of this strand of literature consists in \citet{chung}, who investigate the relations between various stock indices in the NYSE and the AMEX. Ther results suggest that the NYSE is more liquid than the AMEX. The superiority of the NYSE in terms of liquidity is established by showing that the lagged NYSE index has the strongest explanatory power for changes in the price indices of the AMEX. This is obtained by studying the share of price variance of each index explained by exogenous shocks to the NYSE.

\citet{ha02} introduces two additional applications of the VAR methodology. For instance, Hasbrouck (1988, 1991 and 1993) considers the deviations of actual transaction prices from unobservable equilibrium prices. This modelling approach starts with the decomposition of transaction prices into a random walk and a stationary component, along the same lines proposed in macroeconomics by \citet{phillips}. The random walk component identifies the efficient price. The stationary component pins down the difference between the efficient price and the actual transaction price, also called pricing error. The dispersion of the pricing error is a natural measure of market
quality.

The methodology proposed by \citet{ha02} can be used to study a second relevant issue, namely the extent to which fluctuations in a given market arise from swings in another market. This issue has opened the door to the use of cointegration analysis in empirical market microstructure.

\subsubsection{Cointegration in market microstructure}

Financial markets are characterized by `price multiplicity'. In particular, different investors can provide different valuations and attach different prices to the same asset. Also, different market venues can be availbale for the same asset.  Following \citet{ha02}, we can guess a statistical model for the joint behavior of two prices for the same asset linked together by a no-arbitrage or equilibrium relationship. Basically, the two prices incorporate a single long-term
component that takes the form of a cointegrating relation.

Cointegration involves restrictions stronger than those implied by correlation. Two stock prices can be positively correlated but not cointegrated. If stock A is cointegrated with stock B, there exists an arbitrage relationship that ties together the two stocks. In addition, the ask and bid quotes for stock A are also cointegrated. The reason is that the difference between the quotes can often be characterized as a stationary variable, meaning that it cannot explode in an unbounded way. The price of a stock on two different exchanges can be different at any point in time, but it is natural to assume that this difference reverts to its mean over time. 

To provide a simple example, let us consider a security that trades in two different markets. The price on market 1 is denoted as $P_{1t}$, while the price on market 2 is given by $P_{2t}$. Suppose that the two prices are driven by the same efficient price as:
\[
p_{t}=\left[
\begin{array}{c}
p_{1t} \\
p_{2t}%
\end{array}
\right] =\left[
\begin{array}{c}
1 \\
1%
\end{array}
\right] V_{t}+\left[
\begin{array}{c}
S_{1t} \\
S_{2t}%
\end{array}
\right]
\]
where $V_{t}$ is the (unobservable) efficient price, and $S_{1t}$ and $S_{2t}$ are the pricing errors associated to each security. The unobservable efficient price follows a random walk:
\[
V_{t}=V_{t-1}+u_{t}
\]
The component of the pricing error vector can be viewed as originating from bid-ask bounce, price discreteness or inventory effects. In other words, we require that the characteristics of the pricing errors do not generate a permanent effects on prices. This idea can be formalized by assuming that $S_{1t}$ and $S_{2t}$ evolve according to a zero mean-covariance stationary process. From these assumptions, we can obtain a moving-average representation for the return $\Delta p_{t}:$%
\[
\Delta p_{t}=\varepsilon _{t}+\psi _{1}\varepsilon _{t-1}+\psi
_{2}\varepsilon _{t-2}+...
\]
where $\varepsilon _{t}=\left[ \varepsilon _{1t},\varepsilon _{2t}\right]$ consists of innovations reflecting information in the two separate markets. The sum $\Psi(1)=I+\psi _{1}+\psi _{2}+...$, with $I$ as a $2\times 2$ identity matrix, reflects the impact of an initial disturbance on the long-term component. The random-walk variance is:
\[
\sigma _{u}^{2}=\Psi \Omega \Psi ^{^{\prime }}
\]
where $\Omega =Var\left( \varepsilon \right) $. 

The simplest approach for achieving identification in the impat of innovations considers the random-walk variance contribution from both markets:
\[
\sigma _{u}^{2}=\left[
\begin{array}{cc}
\Psi _{1} & \Psi _{2}%
\end{array}
\right] \left[
\begin{array}{cc}
\sigma _{1}^{2} & \sigma _{12} \\
\sigma _{12} & \sigma _{2}^{2}%
\end{array}
\right] \left[
\begin{array}{c}
\Psi _{1} \\
\Psi _{2}%
\end{array}
\right]
\]
If this covariance matrix is diagonal, then we can identify the model in order to deliver a clean decomposition of the random walk variance between the two markets. However, if the covariance matrix is not diagonal, then the covariation between the two prices cannot be easily attributed to either market. \citet{ha96} introduces a bound for the information shares coming from each market through an orthogonalization of the covariance matrix. 

The following question is whether there are alternative restrictions that deliver a better identification. A fruitful approach has been introduced by \citet{hmiw} who assume a generic stochastic process in place for $V_t$. This can be denoted as $f_t$ and assumed I(1), although not necessarily a random walk. Thus, to price in multiple markets, \citet{hmiw} specify a process for $V_t$ such that $V_t=A_{p_t}$, where $A=[a_1, a_2]$ is subject to a normalization $a_1+a_2=1$. The interpretation of the parameter vector $A$ is a very appealing feature of this framework. 

The approach of \citet{mw} suffers of several shortcomings. For instance, it is unclear why one should not consider stochastic processes where past prices reveal additional information. In general, we can modify this simple model to take into account the different effects of revealing both public and private information. \citet{ha02} follows this avenue to to study alternative mechanisms of attribution of price discovery in multiple markets. In this case, the price component of interest is not forced to be a random walk. \citet{ha02} follows this approach because the application of permanent-transitory decompositions to microstructure price data tends to characterize non-martingale pricing factors, which are inefficient proxies for optimally-formed and updated expectations. 

An additional extension in multi-market analysis consists in using price data with differing frequencies. \citet{ha02} shows that the usual method of collecting data from different markets in which trades occur simultaneously can beprovide misleading inferences on price discovery. In this case, information on price leadership may not be accounted for.

\section{Measures based on transaction costs}

Among the transaction costs measures, the bid-ask spread and its variants are the indicators of market liquidity that are used most commonly. The reason is that they convey insight on information sharing in the market. The intuition behind the use of the bid-ask spread lies in the fact that market prices depend the side of the market that initiates the trade. Buyer-initiated trades are concluded at the ask price, while seller-initiated trade are concluded at the bid price. The difference between the best (lowest) ask price and the best (highest) bid price defines the bid-ask spread.

\subsection{The bid-ask spread}

In general, the bid-ask spread is a measure of transaction costs in dealer
markets like the NASDAQ. A market bid is the highest price at
which a dealer is willing to buy a stock, and at which an investor intends to sell. A
market ask is the lowest price at which the dealer is willing to sell the
stock. We should stress that the expression `highest price' stands for `the best market offer'. Since the dealer posts both the bid and ask quotes, the spread between these quantities can be interpreted as the price that the market pays for the liquidity services offered by the dealer. 

\citet{hs96} suggest that specialists often operate as dealers. This is due to the
institutional characteristics of specialists. Typically the specialist disseminates a quote in the market. Market orders are then worked out against limit orders previously placed on the quote posted by the specialist. The disseminated quote is set exactly as the bid-ask spread on the dealer market. 

Let $A_{t}$ denote the ask price, $B_{t}$ the bid price, and $S_{t}$ the
spread at time $t$. Formally, the quoted absolute bid-ask spread is
defined as:
\begin{equation}
S_{t}=A_{t}-B_{t}  \label{sp}
\end{equation}
Frequently the literature reports also a measure of half of the spread, given
by $S_{t}/2$, and the midpoint quote, given by the best bid and ask quotes
in effect for a transaction at time $t$. From (\ref{sp}), we can see that more liquid markets generate lower quoted spreads. This highlights the existence of a negative relationship between the spread and asset prices, as explicitly discussed by \citet{am91}. 

From this simple measure, it is possible to construct additional
indices that are often used to model market liquidity. One of these
consists in the percentage term at which spread is computed. Given the
quote midpoint as $M_{t}=\left( A_{t}+B_{t}\right) /2$, a measure of the
percentage spread $pS_{t}$ is given by:
\begin{equation}
pS_{t}=\frac{A_{t}-B_{t}}{M_{t}}  \label{pers}
\end{equation}
The spread itself represents a measure of transaction costs, rather than a liquidity index in pure sense. However, in a modern market, high transaction costs represent a source of a low liquidity. 

\citet{Cohen} characterize the distinction between the dealer spread and the market spread. The dealer spread is the simple bid-ask spread defined in (\ref{sp}). The market spread, instead, is the difference between the highest bid and the lowest ask across dealers quoting the same stock at the same time. According to \citet{hami2}, a market spread can be lower than a dealer spread. In fact, the cost of immediacy to investors is represented by the size of the market spread. The state of competition and the order processing costs are rather related to the absolute magnitude of the spread. 

An additional issue characterizing the simple spread analysis has to do with the fact that it cannot capture the impact of large block-size transactions on market prices. In fact, the spread measure implicitly assumes that trades occur only at the posted quotes. In this case it is difficult to establish if the transactions occurred for a given price are formed inside or outside the quoted spread. Moreover, given the ability of the market spread to drift away from the one consistent with the perfect market hypothesis, the size of the spread can reflect three main microstructural phenomena. These consist of a pure execution cost, an inventory cost position of the dealer, and an information component cost. In order to detect these various components, the literature has proposed many empirical tests and theoretical models. We review these frameworks in the following sections.

The study of the spread in absolute terms represents only a preliminary stage towards a deeper analysis of transaction costs and of information asymmetry. In fact, the decomposition of the spread components can allow to disentangle the most important effects arising from trading activities.

\subsection{A measure of implied spread}

The measure of \citet{Roll84} is one of the most famous liquidity indices proposed in the microstructure literature. Roll's idea consists in using a model to infer the realized spread (the effective spread) that is reflected from the time series properties of observed market prices and/or returns. 

The main drawback of this type of model is that it does not offer any insights on the possible components of the spread. The reason is that this framework is based on the assumption of homogeneous information across traders. Therefore, the adverse selection component is missing. The magnitude of the spread reflects only the so-called order processing costs, which are considered as having transitory effect, in contrast to information effects, that have permanent effects. 

Let $P_{t}$ denote the observed transaction price of a given asset at time $t$, oscillating between bid and ask quotes that depend on the side originating the trade. We assume that this reproduces the negative serial covariance observed in actual price changes, documented by \citet{famafrench}. The equilibrium price $V_{t}$ follows a pure random-walk process with drift:
\begin{equation}
V_{t}=\overline{V}+V_{t-1}+\varepsilon _{t}  \label{tv}
\end{equation}
where $\varepsilon _{t}$ is the unobservable innovation in the true value of
the asset between transaction $t-1$ and $t$. This is an i.i.d. term with zero mean and
constant variance $\sigma _{\varepsilon }^{2}$. The observed price can be
described as follows:
\begin{equation}
P_{t}=V_{t}+\frac{S}{2}Q_{t}  \label{pri}
\end{equation}
where $S$ denote the quoted absolute spread, assumed to be constant over time. $Q_{t}$ is an indicator function that takes values -1 or 1 with equal probabilities depending on the fact that the $t$-th transaction may occur at the bid or at the ask.\footnote{%
In particular, $Q_{t}=-1$, for seller inititiated transaction, $Q_{t}=+1$,
for transaction initiated by a buyer at the ask quote.} Thus, the
change in transaction prices is given by:
\begin{equation}
\Delta P_{t}=\overline{V}+\frac{S}{2}\Delta Q_{t}+\varepsilon _{t}
\label{deltap}
\end{equation}
To obtain a reduced form, we need two additional assumptions: 
\begin{itemize}
\item the market is
informationly efficient, that is $cov(\varepsilon_{t},\varepsilon _{t-1})=0$, and 
\item buy and sell orders have equal probability, i.e. $cov(\Delta P_{t},\Delta P_{t-1})=-1$. 
\end{itemize}

The probability distribution of trade direction can be represented in the following form:
\begin{center}
\begin{tabular}{ccccccc}
&  &  & \multicolumn{2}{c}{$Q_{t-1}=-1$} & \multicolumn{2}{c}{$Q_{t-1}=+1 $} \\
&  &  & \multicolumn{2}{c}{Trade at bid} & \multicolumn{2}{c}{Trade at ask} \\
\multicolumn{2}{c}{Trade Sequence} & $\rightarrow $ & $B_{t-1}B_{t}$ & $%
B_{t-1}A_{t}$ & $A_{t-1}A_{t}$ & $A_{t-1}B_{t}$ \\
& $\downarrow $ &  & 0 & 2 & 0 & -2 \\ 
& $B_{t}B_{t+1}$ & 0 & 1/4 & 0 &  & 1/4 \\
& $B_{t}A_{t+1}$ & 2 & 1/4 & 0 & 0 & 1/4 \\
& $A_{t}A_{t+1}$ & 0 & 0 & 1/4 & 1/4 & 0 \\
& $A_{t}B_{t+1}$ & -2 & 0 & 1/4 & 1/4 & 0 \\ 
\end{tabular}
\end{center}
Thus, since buy or sell transactions are equally likely, the joint distribution
can be characterized as:
\begin{center}
\begin{tabular}{ccccc}
&  & \multicolumn{3}{c}{$\Delta Q_{t}$} \\
&  & 2 & 0 & -2 \\ 
& 2 & 0 & 0 & 1/8 \\
$\Delta Q_{t+1}$ & 0 & 1/8 & 1/4 & 1/8 \\
& -2 & 1/8 & 1/8 & 0 \\ 
\end{tabular}
\end{center}
It is not difficult to verify that the autocovariance of trades is
given by:
\[
Cov\left( \Delta Q_{t},\Delta Q_{t+1}\right) =-4\cdot \frac{1}{8}-4\cdot
\frac{1}{8}=-1
\]
Therefore, the autocovariance function of price variations is:
\[
Cov\left( \Delta P_{t},\Delta P_{t-1}\right) =Cov\left( \frac{S}{2}\Delta
Q_{t},\frac{S}{2}\Delta Q_{t-1}\right) =\frac{S^{2}}{4}Cov\left( \Delta
Q_{t},\Delta Q_{t-1}\right)
\]
from which we obtain:
\begin{equation}
Cov\left( \Delta P_{t},\Delta P_{t-1}\right) =-\frac{S^{2}}{4}  \label{roll}
\end{equation}

Equation (\ref{roll}) provides the measure of spread defined by Roll (1984). Roll's estimator is obtained by estimating the autocovariance and solving for $S$. The estimator for the serial covariance is:
\begin{equation}
\widehat{Cov}=\frac{1}{n}\sum_{t=1}^{n}\Delta P_{t}\Delta P_{t-1}-\overline{%
\Delta P}^{2}  \label{rolle}
\end{equation}
where $\overline{\Delta P}^{2}$is the sample mean of $\left\{ \Delta P\right\} $. It is possible to show that the population distribution of $\widehat{Cov}$ is asymptotically normal as $n$
increases \citep[see][]{harris}. Moreover, the serial covariance estimator has a downward bias in small samples of data with low frequency. In particular, the bias is large for data with frequency higher than daily.

The implications from (\ref{roll})-(\ref{rolle}) are that the more negative the return autocorrelation is, the higher the illiquidity of a given stock will be. Also, as discussed by \citet{lmk}, there is a relation between the variance ratio and Roll's (1984) measure of liquidity. This link arises from the dependence of the variance ratio on the autocorrelation of daily returns. With returns that exhibit a negative autocorrelation, the measure of \citet{Roll84} generates higher illiquidity and a variance ratio lower than one.

The main shortcoming of this measure consists in its inability to capture asymmetric information effects. The magnitude of the spread described here can be used only to study the size of pure order-processing costs. As stressed by \citet{hs97}, short-term returns can be affected by a multiplicity of factors different from those described here. The point is that the measure of \citet{Roll84} can be safely applied only under the assumption that the quotes do not change in response to trades. This condition would hold only if there were no informed traders in the market, and the quotes did not adjust to compensate for changes in inventory positions.

According to \citet{hs96}, in the case of the NYSE, Roll's measure is much lower than the effective half-spread, while for NASDAQ it is `virtually' identical to the effective spread. This is the same as saying that specialist dealers adjust their quotes in response to the trades because of information effects. At the same time, NASDAQ dealers do not adjust their quotes, thus supporting the assumption of a minimal role for asymmetric information in this market.

\subsection{The role of asymmetric information}

\citet{glo} is the first contribution that models the role of information asymmetries in market microstructure. This paper introduces the distinction between the effects arising from order processing and those from adverse information. As previously remarked, the first type is transitory, while the latter is permanent. On the other hand, the adverse-information component produces non-transitory impacts because it affects the equilibrium value of the security. There are many reasons for price effects to be long-lasting. For instance, this can arise when market-makers engage in trades with investors who possess superior information. Thus, an order placed by a trader can be correlated with the true value of the asset. 

The model of \citet{glo} includes two basic equations:
\begin{equation}
V_{t}=\overline{V}+V_{t-1}+\left( 1-\gamma \right) \frac{S}{2}+\varepsilon
_{t}  \label{tvg}
\end{equation}
\begin{equation}
P_{t}=V_{t}+\gamma \frac{S}{2}Q_{t}  \label{pg}
\end{equation}
where $\gamma $ is the fraction of the quoted spread due to order processing costs, and $\left( 1-\gamma \right) $ is the share arising from adverse information.\footnote{The notation for the other variables is the same as the one outlined in the previous sections.} Note that $\varepsilon _{t}$ reflects the effect arising from the arrival of public information. Thus, the true price $V_{t}$ fully reflects all the information available to the public immediately after a transaction of sign $t$, and the information revealed by a single transaction through the sign of the variable $Q_{t}$. We should stress that the model of \citet{Roll84} is nested by this specification and obtains from $\gamma =1$. It is not difficult to prove that the autocovariance of the price change is equal to:
\begin{equation}
Cov\left( \Delta P_{t},\Delta P_{t-1}\right) =-\gamma \frac{S^{2}}{4}
\label{glosten}
\end{equation}

\subsection{The relation between inventory and adverse-information effects}

Inventory and information effects are key determinants of liquidity conditions. With information effects, prices move against the dealer after a trade. They fall after a dealer purchase, and they rise after a dealer sale. This is often denoted as a `price reversal', and consists of a situation where a dealer trades against informed agents. In this case, a market maker can incur in significant losses.

The idea of price reversal arises from the observation that the realized spread is often different from the quoted spread. In \citet{stoll}, the quoted spread $S$ is taken as constant and depends only on the transaction size, which is constant as well. In practice, the model of \citet{Roll89} assumes that transactions occur either only at the bid or ask quotes. If inventory-holding costs are included into the model, the dealer will have the incentives to change the spread to either  induce or inhibit additional trading movements. In fact, after a dealer purchase (a market sale), bid and ask quotes drop in order to induce dealer sales and disincentivate
additional dealer purchases. However, bid and ask quotes increase after a dealer sale (a market purchase) to inhibit additional dealer sales. 

This type of spread revision operates in the same way both in the case of inventory control and adverse information. However, the reasons for spread revisions are different. With asymmetric information, a buyer (seller) initiated transaction conveys informations on a higher (lower)  expected price of the asset. This is due to the expectation by market participants that active  traders possess superior information. 

Summing up, different reasons for a spread revision can produce similar observed effects. The inventory effect pushes the dealer towards a quote revision in order to avoid a process of trade that would even out his inventory position. With asymmetric information there is the need for the dealer to protect himself from adverse trading directions generated by better informed counterparties.

\subsection{The model of \citet{stoll}}

The model of \citet{stoll} presents a way to jointly estimate the three key components of the spread, namely the shares due to order processing, inventory and adverse information. Notably, this framework allows for the possibility that order flows arising from different motives need not occur with the same probability. To fix the ideas, let $\theta$ denote the probability of a price reversal, i.e. the unconditional probability of a trade change: $\theta =\Pr \left\{Q_{t}=Q_{t-1}\right\}$. The size of a price change conditional on a reversal is given by $\left( 1-\lambda \right) S$. In other words:
\[
\left( 1-\lambda \right) S=\Delta P_{t}\mid \left\{ Q_{t}\neq
Q_{t-1}\right\}
\]
where
\[
\left\{ Q_{t}\neq Q_{t-1}\right\} =\left\{
\begin{array}{c}
P_{t-1}=B_{t-1},P_{t}=A_{t},\text{ or} \\
P_{t-1}=A_{t-1},P_{t}=B_{t}%
\end{array}
\right.
\]
In the framework of \citet{stoll}, the price change $\Delta P_{t}$ arises from the fact that the initial trade is at the bid or at the ask. Thus, for transactions starting at the ask price, we have:
\[
\Delta P_{t}=\left\{
\begin{array}{c}
\left( B_{t}-A_{t-1}\right) =\left( 1-\lambda \right) S\text{, with
probability }\theta \\
\left( A_{t}-A_{t-1}\right) =-\lambda S\text{, with probability }\left(
1-\theta \right)%
\end{array}
\right.
\]
and for transactions starting at the bid price:
\[
\Delta P_{t}=\left\{
\begin{array}{c}
\left( A_{t}-B_{t-1}\right) =\left( 1-\lambda \right) S\text{, with
probability }\theta \\
\left( B_{t}-B_{t-1}\right) =-\lambda S\text{, with probability }\left(
1-\theta \right)%
\end{array}
\right.
\]
Therefore, the expected price change conditional on an initial transaction
at the ask is given by:
\begin{equation}
E\left\{ \Delta P_{t}\mid P_{t-1}=A_{t-1}\right\} =-\left( \theta -\lambda
\right) S  \label{ask}
\end{equation}
while the expected price change conditional on an initial transaction at the
bid is:
\begin{equation}
E\left\{ \Delta P_{t}\mid P_{t-1}=B_{t-1}\right\} =\left( \theta -\lambda
\right) S  \label{bid}
\end{equation}

The realized spread is the dealer's gain after two transactions, consisting of a purchase and a sale. In particular, it denotes the difference between the expected price change after a dealer purchase and the expected price change after a dealer sale. Given the effective spread  $s$, we have:
\begin{equation}
s=2\left( \theta -\lambda \right) S  \label{stoll}
\end{equation}
Note that the realized spread is the remuneration for the services provided by a market maker, including all the components discussed earlier. The fraction of spread given by (\ref{stoll}) includes both the order processing and the inventory component. The adverse information term consists in the fraction of the spread not earned by the market maker, and is equal to: $\left[ 1-2\left( \theta -\lambda \right) S\right] $. 

To provide an empirical implementation of this approach, we can distinguish between two cases, occurring when it is possible to observe directly the trade direction, and when trade data are not available. If market data are available, we can directly
estimate a version of equations (\ref{ask})-(\ref{bid}) in the following form:
\begin{equation}
\left( s_{\tau }^{i}\mid B_{\tau }^{i}\right) =\left[ \left( P_{t+\tau
}^{i}-P_{t}^{i}\right) \mid P_{t}^{i}=B_{\tau }^{i}\right]  \label{ebid}
\end{equation}
for trades at the bid and:
\begin{equation}
\left( s_{\tau }^{i}\mid A_{\tau }^{i}\right) =\left[ \left( P_{t+\tau
}^{i}-P_{t}^{i}\right) \mid P_{t}^{i}=A_{\tau }^{i}\right]  \label{eask}
\end{equation}
for trades at the ask. In equations $\left( \ref{ebid}\right) $-$\left(\ref{eask}\right) $, $\tau $ indicates the time length after which a subsequent price is observed. The choice of the time horizon adopted in the estimation is crucial. If the time frame is too short, the subsequent price may fail to reflect a reversal, and may reflect only another trade in
the same direction. However, if the time horizon is too long, we might obtain results affected by excessive price volatility due to frequent conseutive price changes. 

\citet{hs96} run an empirical exercise by using four alternative time horizons, namely  between five and ten minutes after the initial trade at $t$, with the first trade occurring at least five minutes after the initial trade, with the first trade between 30 and 35 minutes after the initial trade, and with the first trade occurring at least 30 minutes after the initial trade. The findings of \citet{hs96} suggest that dealers in the NASDAQ face a lower realized spread than on NYSE.

When trade data are not available, we need to resort to information from the statistical patterns characterizing an asset price. It can be shown that the covariance of price changes is given by:
\[
Cov\left( \Delta P_{t},\Delta P_{t+1}\right) =S^{2}\left[ \lambda ^{2}\left(
1-2\theta \right) -\theta ^{2}\left( 1-2\lambda \right) \right]
\]
In order to detect inventory costs, \citet{stoll} presents also the autocovariance of changes in quotations. This takes the form:
\[
Cov\left( \Delta Q_{t},\Delta Q_{t+1}\right) =S^{2}\lambda ^{2}\left(
1-2\theta \right) \qquad \qquad Q=A,B
\]
Under the assumption of constant quoted spread, this covariance can be computed either from changes in the bid or the ask quotes, so that $Cov\left( \Delta B_{t},\Delta B_{t+1}\right) =S^{2}\lambda ^{2}\left( 1-2\theta \right) $, or $Cov\left( \Delta A_{t},\Delta A_{t+1}\right) =S^{2}\lambda ^{2}\left(1-2\theta \right) $. The expressions for the covariance delivered by different versions of the model are collected in table \ref{tabs1}.

\begin{table}[!t]\label{tabs1}
\begin{center}
\caption{Covariance in the model of \citet{stoll}}
\begin{tabular}{lcc}
\hline\hline
Spread Determinant & $Cov\left( \Delta P_{t},\Delta P_{t+1}\right) $ & $%
Cov\left( \Delta Q_{t},\Delta Q_{t+1}\right) $ \\ \hline
Order Processing: $\theta =1/2,$ $\lambda =0$ & $-\frac{1}{4}S^{2}$ & 0 \\
Adverse Information: $\theta =1/2,$ $\lambda =0.5$ & 0 & 0 \\
Inventory Costs: $\theta >1/2,$ $\lambda =0.5$ & $\left( -\frac{1}{4}%
S^{2},0\right) $ & $\left( -\frac{1}{4}S^{2},0\right) $ \\ \hline
\end{tabular}
\end{center}
\end{table}

The theory of bid-ask spread considered thus far is based on several assumptions that can be challenged. \citet{gkn} show that the available estimators of spread components are typically biased and inefficient. This is due to two important facts. The first one is that stock returns contain a statistically significant component that is positively autocorrelated, as showed by \citet{gkn}. Moreover, transaction returns display a large unexpected return component. \citet{gkn} introduce time-varying expected returns by assuming $\overline{V}_{t}\neq \overline{V}$. According to their model, the autocovariance of quote changes are positive. 


Other approaches for the estimation of the spread components include the so-called `trading indicators'. The models proposed in this context, such as \citet{glha}, \citet{glo} and \citet{madha} do not contain any assumption about the arrival of orders. Only the actual direction of trades affects the parameter estimation. 

Given their structure, these models require a very careful specification of the type of market under study. In this respect, they are not general enough. Depending on whether there is a quote or an order driven market, we can observe a different behavior of the transaction price that is related to the order size. Therefore, for small transactions, the model of \citet{glha} underestimates the adverse selection component, and overestimates the order processing component. The opposite holds for transactions of large blocks. Within the class of models of trade indicators, the framework proposed by \citet{madha} allows to disentangle the effects from adverse information and inventory changes. The price mechanism proposed displays an asymmetric information component of the spread related to innovations in the order flow. \citet{madha} construct a trade indicator model that allows to consider also trades inside the quotes. The model proposed is general enough to capture the information advantages linked to unexpected trading movements.

\subsubsection{The model of \citet{hs97}}

The framework proposed by \citet{hs97} introduces a three-part decomposition of the spread. The model is based on the following unobservable equilibrium price as:
\begin{equation}
V_{t}=V_{t-1}+\eta \frac{S}{2}Q_{t-1}+\varepsilon _{t}  \label{hs1}
\end{equation}
In equation (\ref{hs1}), the term $\eta $ denotes the percentage of the half spread due to adverse selection, $\varepsilon _{t}$ is a public information shock serially uncorrelated over time. $Q_{t}$ is a trade indicator variable, which is equal to 1 if the transaction is buyer-initiated, and is equal to -1 if the transaction is started at the bid. 

The last trade conveys relevant information in determining the true value of the stock price. Given a midpoint quote $M_{t}=\left( A_{t}+B_{t}\right) /2$, \citet{hs94} assume the following relationship with the unobserved price:
\begin{equation}
M_{t}=V_{t}+\delta \frac{S}{2}\sum_{i=1}^{t-1}Q_{i}  \label{hs2}
\end{equation}
where $\delta $ measures the inventory effect, and $\sum_{i=1}^{t-1}Q_{i}$ is
the cumulative inventory from market opening until $t-1$. In particular, $%
Q_{1}$ is the initial inventory of the day. Combining (\ref{hs1}) with (%
\ref{hs2}), we obtain the change in the midpoint quote:
\begin{equation}
\Delta M_{t}=\left( \delta +\eta \right) \frac{S}{2}Q_{t-1}+\varepsilon _{t}
\label{hs3}
\end{equation}
The traded price $P_{t}$ is:
\begin{equation}
P_{t}=M_{t}+\frac{S}{2}Q_{t}+u_{t}  \label{hs4}
\end{equation}
In this model $S$ denotes the traded spread. This is different from the quoted
(posted) spread $S_{t}$ because it reflects also trades inside the
quotes but outside the midpoint. Combining (\ref{hs1})-(\ref{hs4}) we obtaint:
\begin{equation}
\Delta P_{t}=\frac{S}{2}\Delta Q_{t}+\left( \delta +\eta \right) \frac{S}{2}%
Q_{t-1}+\zeta _{t}  \label{hs5}
\end{equation}
where $\zeta _{t}=\Delta u_{t}+\varepsilon _{t}$. Equation (\ref{hs5})
reflects only a two-way decomposition of the spread. The order processing
cost is defined as $1-\delta -\eta$. However, by estimating equation (\ref%
{hs5}) alone, it is not possible to draw any conclusion on either the
relative importance of the adverse-information component, or the inventory effect. Only a simultaneous three-way decomposition of the spread can fully uncover all these effects jointly. For this purpose, we need to add to the model an additional equation specifying the probability of trade direction.

\citet{hs97} introduce the assumption of serial correlation in trade
flows:
\begin{equation}
E\left( Q_{t-1}\mid Q_{t-2}\right) =\left( 1-2\theta \right) Q_{t-2}
\label{hs6}
\end{equation}
with $\theta $ as the probability of change in trade direction (i.e. the
probability that a trade at the bid at time $t-1$, is followed by a trade at
the ask at time $t$). For $\theta \neq 0.5$, the change in the true price is given by:
\begin{equation}
\Delta V_{t}=\eta \frac{S}{2}Q_{t-1}-\eta \frac{S}{2}\left( 1-2\theta
\right) Q_{t-2}+\varepsilon _{t}  \label{hs7}
\end{equation}
Equation (\ref{hs7}) has three components. The first one, represented by $\eta\frac{S}{2}Q_{t-1}$,  displays the information conveyed by the last trade. The second part, given by $\eta \frac{S}{2}%
\left( 1-2\theta \right) Q_{t-2}$, introduces the additional persistence in information that is not accounted for by the surprise term $\varepsilon _{t}$. If $\theta =1/2$ equation (\ref{hs7}) collapses into (\ref{hs1}). The reader should note from (\ref{hs6}) and (\ref{hs7}) that changes in the true value of the asset are unpredictable until the release of public information contained in $\varepsilon _{t}$ shows up, so that $E\left( \Delta V_{t}\mid V_{t-1},Q_{t-2}\right) =0$. We can combine (\ref{hs7}) and (\ref{hs2}) to obtain the change in midpoint quote:
\begin{equation}
\Delta M_{t}=\left( \delta +\eta \right) \frac{S}{2}Q_{t-1}-\eta \frac{S}{2}%
\left( 1-2\theta \right) Q_{t-2}+\varepsilon _{t}  \label{hs8}
\end{equation}
Equation (\ref{hs8}) stresses the fact that the inventory effect can be detected only after the trades are executed. In this case, the quotes are revised. This allows to distinguish
between the adverse-information component and the inventory component. 

Taking the expectation of equation (\ref{hs7}) conditional on the information obtained after observing $M_{t-1}$ and before $Q_{t-1}$ and $M_t$, we get:
\begin{equation}
E\left( \Delta M_{t}\mid M_{t-1},Q_{t-2}\right) =\delta \frac{S}{2}\left(
1-2\theta \right) Q_{t-2}  \label{hs9}
\end{equation}
From this equation we can see that the expected change in the midpoint depends only on $\delta $, the inventory cost component. The inventory-quote response to a trade is given by $\delta \frac{S}{2}$. However, from (\ref{hs8}), the change in the midpoint quote due to the inventory effect is much smaller. To get the three-way spread decomposition, we can combine (\ref{hs4}) and (\ref{hs8}) to deliver:
\begin{equation}
\Delta P_{t}=\frac{S}{2}Q_{t}+\left( \delta +\eta -1\right) \frac{S}{2}%
Q_{t-1}-\eta \frac{S}{2}\left( 1-2\theta \right) Q_{t-2}+\zeta _{t}
\label{hs10}
\end{equation}
Thus, by estimating simultaneously (\ref{hs6})-(\ref{hs10}), we can identify all the three spread components, namely $\delta $, $\eta $ and $1-\delta-\eta $, together with the probability of a trade reversal $\theta $. The reader should note that $S$ denotes the effective spread, which is estimated. If the traded spread is replaced with the posted spread $S_{t}$, then model consists of equation (\ref{hs6}) and:
\begin{equation}
\Delta M_{t}=\left( \delta +\eta \right) \frac{S_{t-1}}{2}Q_{t-1}-\eta \frac{%
S_{t-2}}{2}\left( 1-2\theta \right) Q_{t-2}+\varepsilon _{t}  \label{hs11}
\end{equation}
In this case, the parameter space is reduced. This is beneficial only when a limited dataset is available.

\citet{hs97} test the model on trades and quotes for large and actively-traded stocks in 1992. From their results, the average order processing of the traded spread is 61.8\%, the average adverse-information component is 9.6\%, and the average inventory cost component is 28.7\%. Another interesting piece of evidence consists in the fact that the adverse information component of the spread is smaller for large trades. This is due to the fact that large trades usually tend to be negotiated outside the market, so that the price fully reflects the information given by the last trade.

\subsubsection{Empirical issues}

An integrated approach on the analysis of the spread has been proposed by \citet{hs94}. They consider a two-equation framework where the determinants of quotes and transaction prices are included to test for the relevance of competing microstructure theories. 

To shed light on the issue, let us consider the logarithm $M_t$ of the midpoint quote. The price (in logs) can be expressed as follows:
\begin{equation}
P_{t}=M_{t}+W_{t}  \label{h1s}
\end{equation}
where $W_{t}$ is the deviation of the log of observed transaction price $P_{t}$ from the log-midpoint quote. This suggests that trades can occur also inside the quotes. Thus, the effective spread is always less than the quoted spread. In equation (\ref{h1s}), public dealer purchases (sales) result in $W_{t}>0$ (<0).

We take the first difference of equation (\ref{h1s}) to get:
\begin{equation}
P_{t}-P_{t-1}=M_{t}-M_{t-1}+W_{t}-W_{t-1}  \label{h2s}
\end{equation}
Let us define the return from official quotes $R_{t}^{p}=P_{t}-P_{t-1}$, with $R_{t}^{m}=M_{t}-M_{t-1}$. In order to take the model to the data, we can specify the quote setting behavior from the midpoint change $R_{t}^{m}$. In doing so, \citet{hs94} identify a fourth microstructure effect that is not captured by previous models. The induced order-arrival effect captures the idea that the probability of a public purchase changes through time after a dealer price adjustment. This is determined by the ability of the market maker to induce changes in order arrivals to cover for the entire cost of processing the orders. In general, the induced order arrival effect can be written as:
\begin{equation}
\Pr \text{ob}\left[ W_{t}>0\mid \left( V_{t}-M_{t}\right) >0\right] >0.5
\label{h3s}
\end{equation}
As stated in (\ref{h3s}), the divergence between the unobservable price $V_{t}$ and the midpoint quote depends on the inventory holdings of the supplier.

The subsequent step considered by \citet{hs94} consists in specifying the
pattern of quote returns $R_{t}^{m}$ as follows:
\begin{equation}
R_{t}^{m}=E\left[ R_{t}^{V}\mid \Omega _{t-1}\right] +g\left( \Delta
I_{t-1}\right) +\varepsilon _{t}  \label{h4s}
\end{equation}
where $E\left[ R_{t}^{V}\mid \Omega _{t-1}\right] $ is the expected value of the consensus return, i.e. the return earned on the expected price changes of the true price (expressed in logs): $R_{t}^{V}=V_{t}-V_{t-1}$. In equation (\ref{h4s}), $\Omega _{t-1}$ denotes the set of information available at time t, while $g\left( \Delta I_{t-1}\right) $ is the inventory change of the quote return. To study the information effects, the expected component in equation (\ref{h4s}) can be conditioned on a subset of variables reflecting the availability of public information:
\begin{equation}
E\left[ R_{t}^{V}\mid \Omega _{t-1}\right] =f\left(
W_{t-1},R_{t-1}^{F}\right)  \label{h5s}
\end{equation}

In \citet{hs94}, the term $R_{t-1}^{F}$ denotes the change in logarithm of the S\&P500 futures price. The presence of the term $W_{t-1}$ reflects the adjustment of the market-maker to
public information revealed through trading. If private information was the main source of the bid-ask spread, the quote's midpoint would be adjusted by $W_{t-1}$ because the previous price deviation is the expected value of the private information conveyed by trade. 

The general specification of the model for the return on quote revision is:
\begin{equation}
R_{t}^{m}=\eta _{0}+\eta _{1}R_{t-1}^{m}+\eta _{2}R_{t-1}^{F}+\eta
_{3}W_{t-1}+\eta _{4}H_{t-1}+\eta _{5}L_{t-1}^{A}+\eta _{6}L_{t-1}^{B}+\eta
_{7}Z_{t-1}+\varepsilon _{t}  \label{h6s}
\end{equation}
The inclusion of $R_{t-1}^{F}$ is motivated by the fact that trading in stock index futures is cheaper than trading in stocks. Thus, the diffusion of news can be detected through movements of index futures. \citet{hs94} include $W_{t-1}$ to account for the information from previous period's trading. \citet{hs94} also consider the cumulative volume traded on the single asset, given by $H_{t-1}$. In order to detect the inventory effect, equation (\ref{h6s}) includes two trade indicator variables constructed as follows:
\begin{eqnarray*}
&&L_{t-1}^{A}\left\{
\begin{array}{ccccc}
=1 &  & if\text{ }W_{t-1}>0 & \text{and } & Vol_{t-1}>10,000 \\
=0 &  & \text{otherwise} &  &
\end{array}
\right. \\
&& \\
&&L_{t-1}^{B}\left\{
\begin{array}{ccccc}
=1 &  & if\text{ }W_{t-1}<0 & \text{and } & Vol_{t-1}>10,000 \\
=0 &  & \text{otherwise} &  &
\end{array}
\right.
\end{eqnarray*}
where $Vol_{t-1}$ indicates the share volume traded at time $t-1$ for the asset. The expected impact on the quote revision is positive for $L_{t-1}^{A}$ and negative for $L_{t-1}^{B}$.

Another crucial variable in equation (\ref{h6s}) is represented by the quote revision return $R_{t-1}^{m}$. This allows to take into account non-instantaneous quote revisions, as well as the negative serial correlation in quote returns. Finally, quote returns are affected by the difference between the logarithm of the quoted volume at the ask (depth at the ask) and the logarithm of the quoted volume at the bid (depth at the bid), which is denoted as $Z_{t-1}$. The presence of inventory effects would imply a positive impact on $R_{t}^{m}$. In fact, if a dealer has a large inventory position, he has an incentive to reduce quotes and to raise depth at the ask to encourage transactions with the purpose of mitigating the inventory position. 

In equation (\ref{h6s}) there is also a signalling effect captured by the sign of $\eta _{7}$. If $\eta _{7}<0$, we have a negative impact on quote changes, i.e. a large depth at the ask at time $t-1$ signals the presence of sellers in the limit order book, inducing market participants to revise
quotes downward at time t.

To close the model, \citet{hs94} make an assumption about the stochastic process
for $W_{t}$:
\begin{equation}
W_{t}=\rho W_{t-1}+\xi _{t}  \label{h7s}
\end{equation}
where $\xi _{t}$ denotes the order arrival shock. For $\rho=0$, the probability of a purchase or a sale is independent from the sequence of trades. If the activity of dealer pricing creates an inventory effect, then $\rho <0$. By combining equations (\ref{h3s}) and (\ref{h7s}), we obtain:
\begin{equation}
R_{t}^{p}=R_{t}^{m}+\left( \rho -1\right) W_{t-1}+u_{t}  \label{h7sbis}
\end{equation}
Plugging equation (\ref{h6s}) into (\ref{h7sbis}) delivers the expression for observed returns:
\begin{equation}
R_{t}^{p}=\eta _{0}+\eta _{1}R_{t-1}^{m}+\eta _{2}R_{t-1}^{F}+\eta
_{3}^{p}W_{t-1}+\eta _{4}H_{t-1}+\eta _{5}L_{t-1}^{A}+\eta
_{6}L_{t-1}^{B}+\eta _{7}Z_{t-1}+u_{t}  \label{h8s}
\end{equation}
where $\eta _{3}^{p}=\eta _{3}+\rho -1$, and $u_{t}=\varepsilon _{t}+\xi_{t}$. From this transformation, we see that the coefficient $\eta_{3}^{p} $ can now be decomposed into three components: 
\begin{itemize}
\item[(i)] the asymmetric information effect, given by $\eta _{3}$, which is expected to be positive and represents the information conveyed by the last trade; 
\item[(ii)] the induced order arrival effect, given by $\rho $; 
\item[(iii)] the bid-ask bounce effect. In the absence of information effect, the third component is equal to -1, thus representing the tendency of price returns towards being serial autocorrelation. 
\end{itemize}

In empirical applications, the model consists of the two equations (\ref{h6s}) and (\ref{h8s}). These are jointly estimated by GMM on intraday data. \citet{hs94} report results for the 20 most actively traded stocks in the NYSE.

The specification described in (\ref{h8s}) includes the most important ingredients of the microstructure theory. By setting $\eta _{3}=0$ and $\eta_{3}^{p}=-1 $, we can test the order processing theory of the bid-ask spread. With $\eta _{3}=1,$ $\eta _{3}^{p}=0$, we have the adverse information theory of the spread, where quotes are adjusted in order to
reflect the last trade $Z_{t-1}$. Additionally, the inventory holding cost
theory can be obtained by setting $1>\eta _{3}>0$, $0>\eta _{3}^{p}>-1$ and $\eta _{4}>0$. In this case, the direction of change in quotes follows the last trade. Moreover, quote returns are adjusted by an amount that depends on the inventory change. The induced order-arrival effect theory arises from $\rho <0$, $\eta _{1}<0$ and $\eta _{7}>0$. The key implication of this theory is that changes in the midpoint quotes affect the order arrivals, leading to serial
correlation in $W_{t}$. An important factor is captured by changes in market depth. If
depth is used to encourage order arrival, then we would expect a positive $\eta _{7}$. Alternatively, depth can be a signal or act as a barrier, leading to a negative sign in $\eta _{7}$. The effect of large trades in the adverse information theory is captured in the form of a sign pattern as $\eta _{5}>0 $ and $\eta _{6}<0$. 

Summing up, the two equations (\ref{h6s}) and (\ref{h8s}) produce a full set of testable implications and cross-equation restrictions. Among the various theories of market microstructure that can be tested, we can also include the efficient market hypothesis of index futures. If an asset market is efficient, the predominant prices fully reflect the information contained in the index futures prices. In this case, we would expect $\eta _{2}=0$.

\end{document}